\documentclass[showpacs,showkeys,amsmath,nofootinbib,amssymb,floatfix,twocolumn]{revtex4}
\usepackage[T1]{fontenc}
\usepackage[latin9]{inputenc}
\usepackage{amsmath}
\usepackage{graphicx}

\begin{document}

\title{A new mathematical model for molecular
dynamics 1: \\Molecular basis of memory}

\author{C. K. Raju}

\affiliation{Inmantec, National Highway 24, Dasna Crossing, Ghaziabad, 201 009, Delhi NCR, India}

\altaffiliation{This work was commenced at Centre for Computer Science, MCRP University, Bhopal, India}

\email{c.k.raju@inmantec.edu}

\begin{abstract}
Proteins have been empirically linked to memory. If memory relates to protein structure, then each conformation would \textit{functionally} code only one bit, making it difficult to explain large memories. Nor is there a simple way to relate memory to protein dynamics on current molecular dynamics (MD), which is memoryless. Here we point out that MD may be modified to involve memory \textit{ab initio} without any new hypothesis: simply replace the electrostatic (Coulomb) force by the electrodynamic force---which is more accurate. We now need to solve functional differential equations (FDEs), instead of the ordinary differential equations (ODEs) currently solved in MD. Unlike ODEs, retarded FDEs are history-dependent: so memory is already present even at the level of interacting sites within molecules. The resulting increase in computational complexity is within the reach of current computers. While Amdahl's law does pose a challenge to parallelised time-stepping with this model, the compute-intensive part---the force calculation---may still be carried out in parallel. Thus, reformulating MD to use FDEs is feasible, and this could help to understand the possible dynamical basis of memory.

\end{abstract}
\maketitle

\section{Introduction}

The functioning of biological macromolecules is linked to their structure and dynamics. The determination of structure through X-ray crystallography and NMR is expensive, and these techniques provide no information about the dynamics of protein folding and protein interaction. Molecular dynamics (MD) simulations are typically used for this purpose \cite{Karplus}, and many computer programs have been developed for MD, such as \textsc{amber} \cite{amber}, 
 \textsc{charmm} \cite{charmm}, 
\textsc{gromos} \cite{gromos}, and
\textsc{namd} \cite{namd2}. 

Two key problems are (a) to determine the conformational structure(s) of protein molecules given the sequence of amino acids, and (b) to determine the dynamics of the folding process. The first problem is usually studied by minimizing the free energy.  Solving the Schr\"{o}dinger equation for something as complex as the collection of particles constituting a protein molecule is too far beyond existing computational capabilities, so the second problem has long been studied by solving the Newtonian equations of motion for a prescribed force field involving van der Waals and electrostatic forces. 
 
Some features however seem hard to understand with this approach. First there is the classic Levinthal's paradox: the conformational space to be explored is very vast in relation to the actual folding times that are observed. It has been hypothesised that this is because the energy landscape has a funneled character. While some attempts have been made to derive the funneled landscape from molecular dynamics, none of these is convincing. This hypothetical landscape is made more complex by possible multiple minima, evidence for which comes from prions.

It is even harder to understand how long-term memory formation in biological organisms links to protein molecules. Empirical studies have suggested that the formation of memory correlates with the formation of protein molecules, and preventing protein molecules from forming impairs formation of long-term memories. But if a protein molecule can exist in only a few stable conformations, then a large amount of functionally useful information cannot be stored in its structure, for each conformation corresponds to just one piece of functional information, irrespective of how that form was reached.\footnote{There have been other, unrelated attempts to limit how much information a protein molecule can store, as in \cite{multibit, thermolimit}.}  

Given the short folding times of protein molecules, the transitional state surely cannot be the seat of long-term memory. So, if protein molecules store memory structurally, then multibit information would require multiple conformations---to store memory on the scale of biological organisms, a very large number of conformations would need to be associated with each protein molecule. If protein molecules do indeed admit such a large number of stable conformations, then we are back to a situation reminiscent of Levinthal's paradox, and we need to suppose that the energy landscape has a funnel with numerous minima or perhaps a multi-funnel landscape \cite{multifunnel}. Instead of thus accumulating hypotheses it seems better to look for a simpler solution. 

On the other hand, it is not immediately clear how memory can be present in the dynamics of protein interactions either. Thus, MD  simulations currently solve Newtonian equations of motion which are ordinary differential equations (ODEs). The solution
of a system of ODEs is uniquely determined by the initial data (or
data at any one instant of time), irrespective of past history. Thus, the time evolution of a system modeled by
ODEs must be memoryless or history-independent. Hence,  current MD, which involves memoryless dynamics, cannot readily explain memory-dependent molecular interactions in a simple way. It is possible, of course, that, as in artificial neural networks, memory is holistic and intrinsically beyond the reach of MD which is reductionist.

Before concluding that a simple explanation for memory is intrinsically beyond the reach of molecular dynamics, it seems worthwhile to explore a more sophisticated mathematical model for MD which admits memory \textit{ab initio}. The suggestion to explore such a model is not intended to exclude more holistic models of memory, but to complement them.  Now, given the vast amount of experimental data that is currently available, it is not difficult to invent a variety of \textit{ad hoc} mathematical models which will fit some of it. However, that  would go against the spirit of MD which has the great theoretical virtue that it proceeds directly from established physics, without invoking fresh hypotheses, except as simplifying assumptions.

The modified model for MD, that we propose, needs no additional hypothesis. If existing physics is correctly applied, the
electrodynamic interaction between two moving electrical charges is
history dependent, hence already involves memory. If this feature is incorporated into molecular dynamics at the outset, it may help to understand how protein molecules relate to memory. 

The argument, in outline, is very simple. Currently, the electrostatic (or Coulomb) potential is used in MD for long-range
forces. However, the interacting sites are usually in motion so the Coulomb
force is only an approximation to the full electrodynamic force. If
one uses the full electrodynamic force, the equations to be solved are functional differential
equations (FDEs) \cite{ckrelecmag}. 
Unlike ODEs the solution of these FDEs cannot be uniquely determined
merely by initial data. Assuming these FDEs to be retarded, one must prescribe some part of the past history \cite{elsgoltz,driverbook,ckrelecmag}.
So some rudimentary memory is already present at the level of the
constituents. 

In the case of two interacting elementary charges, a proton and electron, in a hydrogen atom, say, the system memory or the exact time interval over which past data is required is tiny, being of the order of a  deci femto second. However, even in this case the difference
between FDEs and ODEs is significant since solutions of retarded
FDEs may exhibit qualitative behaviour that is \textit{impossible}
for solutions of ODEs.

Further, the system memory, or the time period over which the past history of
the system must be specified, increases with the number of interacting
particles and the scale of the system. Thus, we have a
simple reason to expect more complex molecules and collections of
molecules to exhibit longer-term memory. With currently available computational resources, computing solutions of FDEs for such large collections may be contemplated.

This argument is elaborated below.

\section{Existing and proposed molecular dynamics}

\noindent A wide variety of force fields have been used in molecular
dynamic simulations of biological macromolecules \cite{hobzaetal}.
These force fields are typically derived from a potential represented
as a sum over bonded and non-bonded pairs \cite{mccammon}.
The interaction potential $V_{ij}^{nb}$ between non-bonded pairs
is represented as a sum of a short-range van der Waals potential (usually
a 12--6 Lennard-Jones potential) and the long-range electrostatic
(Coulomb) potential:

\begin{equation}
V_{ij}^{nb}=\left(\frac{A}{r_{ij}^{12}}-\right)+\frac{q_{i}q_{j}}{4\pi\epsilon r_{ij}}\label{stock}\end{equation}

\noindent Here, $A$ and $C$ are constants, $q_{i}$ and $q_{j}$
are charges, while $r_{ij}$ is the distance between them.

Our concern here is with the \textit{second} (electrostatic) term,
on the right hand side of \eqref{stock}. No one earlier seems to
have questioned the validity of using the Coulomb potential---which
is used across various force fields in a variety of computer programs
like \textsc{amber} \cite{amber}, 
 \textsc{charmm} \cite{charmm}, 
\textsc{gromos} \cite{gromos},
\textsc{namd} \cite{namd2},  
etc. currently used for MD simulations.

However, it is elementary that the Coulomb potential applies only
to \textit{static} charges, whereas in molecular dynamics the interacting
charges are in relative motion. In the electrodynamic case, with moving
charges, the force ${\bf F}_{ij}$ on charge $i$ due to charge $j$
is actually obtained \textit{not} from the Coulomb potential but from
the (retarded) Lienard-Wiechert potential and the Heaviside-Lorentz
force law (usually called just the Lorentz law), to give \cite{griffiths3}:

\begin{eqnarray}
{\bf F}_{ij}\: & = & \:\frac{q_{i}q_{j}}{4\pi\epsilon_{0}}\frac{R}{({\bf R\cdot u)^{3}}}\left\{ \left[(c^{2}-v^{2}){\bf u}+{\bf R}\times(u\times a)\right]\right.\\\nonumber
 &  & {}+\frac{{\bf v}_{i}}{c}\times\left.\left[{\bf \hat{R}}\times[(c^{2}-v^{2}){\bf u}+{\bf R}\times({\bf u}\times{\bf a})]\right]\right\} .\label{em-force}\end{eqnarray}

\noindent Here, charge $q_{i}$ is located at ${\bf r}_{i}(t)$ at
time $t$, while the position of the other charge $q_{j}$ at time
$t$ is given by ${\bf r}_{j}(t)$, ${\bf R={\bf r}_{i}(t)-{\bf r}_{j}(t_{r})}$,
and $R$ is its norm. In the preceding expression, $t_{r}$ is the
retarded time (the time at which the backward null cone with vertex
at ${\bf r}_{i}(t)$ meets the world line ${\bf r}_{j}(t)$ of the
other charge), and is given implicitly by the equation $||{\bf r}_{i}(t)-{\bf r}_{j}(t_{r})||=c(t-t_{r})$,
with $||\cdot||$ denoting the 3-vector norm, and $c$ being the speed
of light. Further (with dots denoting time derivatives), ${\bf v}_{i}=\dot{{\bf r}}_{i}(t)$,
${\bf u}=c\frac{{\bf R}}{R}-{\bf v}=c{\bf \hat{R}}-{\bf v}$, and
it is understood that ${\bf v}={\dot{{\bf r}}}_{j}(t_{r})$ and ${\bf a}={\bf \ddot{r}}_{j}(t_{r})$
are the velocity and acceleration of the charge $q_{j}$ at the \textit{retarded
time} $t_{r}$. A similar expression gives the force ${\bf F}_{ji}$
exerted on the charge $q_{j}$ by the charge $q_{i}$. Unlike the
electrostatic case, ${\bf F}_{ji}\neq~{\bf F}_{ij}$, in general.
(Relativistic velocity effects are here ignored as irrelevant to molecular
dynamics, although the conclusions apply \textit{a fortiori} to the
relativistic case.)

The approximation of the full electrodynamic force \eqref{em-force} based on the Lienard-Wiechert potentials, 
by the simpler electrostatic force based on the Coulomb potential \eqref{stock} has a long history, going back to the days of the Rutherford model, and the
Bohr atom. The general \textit{expectation} among physicists has been
that the use of the electrostatic force greatly simplifies calculations, and
would not upset any key qualitative conclusion. Though long-standing,
and widespread, this expectation is mathematically incorrect, and
has never been actually tested. Testing would require that one obtain
the actual solution of at least the 2-particle equations with the
full electrodynamic force \eqref{em-force}, and check whether it
is approximately the same as the solution obtained with the electrostatic
inverse square law force. Prior to the advent of high-speed computers, doing this was well-nigh impossible: the one-particle case \cite{plass}  
being complicated enough, despite various attempts in the previous
century \cite{dirac,synge,schild,wigner1,wigner2,driver,andersonbaeyer},
no actual solutions of the 2-particle equations with the full electrodynamic
force were published, except \cite{hsing} 
in oversimplified situations of little practical interest.

The use of the full electrodynamic force \eqref{em-force} makes a fundamental mathematical difference, for it leads to
the formulation of the $n$-body problem ($n\geq2$) as a system of
FDEs.

\section{FDEs Vs other methods}

To see how FDEs differ fundamentally from ODEs, consider, for example,
the simple FDE \begin{equation}
\frac{dx}{dt}=x(t-\frac{\pi}{2}).\label{egfde}\end{equation}

\begin{figure}[htb]
\includegraphics[width=20pc]{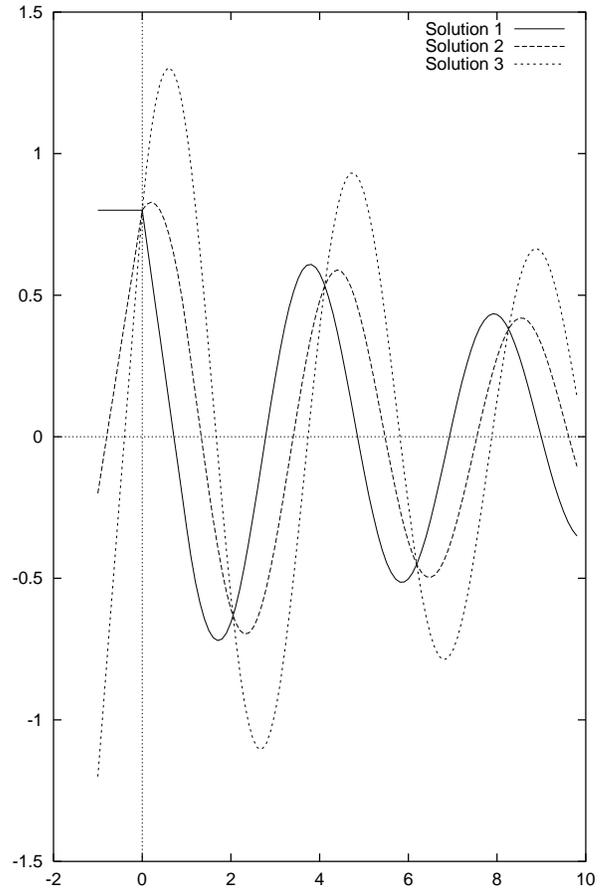} 
\caption{\textbf{Insufficiency of initial data.} The figure shows three different
solutions of a retarded FDE $\dot{x}(t)=x(t-1)$, corresponding to
three different past histories (prescribed for $t\le0$). All solutions have
the same initial data ($x(0)$). The solution of a retarded FDE cannot
be determined uniquely merely by prescribing initial data.}
\label{Figure1} 
\end{figure}

It is easy to verify that both $\cos t$ and $\sin t$ are solutions
of \eqref{egfde}. Since the equation is linear, $a\cos t+b\sin t$
is a solution of \eqref{egfde} for arbitrary constants $a$, $b$,
and it is clear that the values of both constants $a$ and $b$ cannot
be determined by a single initial condition $x(0)=x_{0}$, say. The
situation is shown in Fig.~\ref{Figure1}: there is an infinity of
non-unique solutions if we prescribe the state of the system at only
one instant of time. To obtain a unique solution one must specify
the past history of the system. A system modeled by FDEs, in effect,
has memory.

\begin{figure}[htb]
\includegraphics[width=18pc]{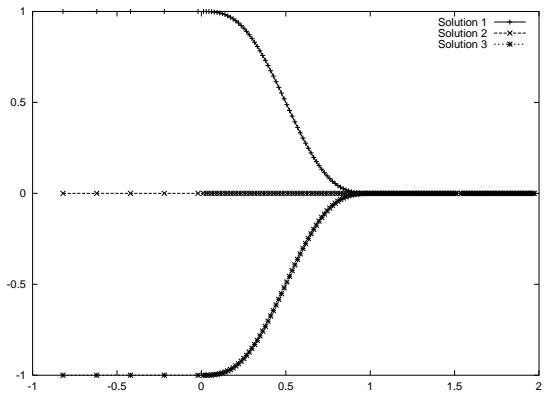}
\caption{\textbf{History Dependence of FDEs.} The figure shows three different
solutions of a retarded FDE $\dot{x}(t)=b(t)x(t-1)$, for a suitable
choice of the function $b(t)$, as described in \cite{ckrelecmag}.
Three different past histories prescribed for $t\leq0$ lead to three
different solutions all of which coincide for $t\geq1$. Such a phase
collapse is impossible with ODEs where trajectories in phase space
can never intersect. Because of this phase collapse. retarded FDEs,
unlike ODEs, cannot, in general, be solved backward, from prescribed
future data.}
\label{Figure2} 
\end{figure}

For the interaction of a proton and an electron within the classical hydrogen atom, the time interval over which past history
must be prescribed is quite small. However, FDE exhibit qualitatively
novel features, such as time asymmetry, and breakdown of phase flow
as illustrated in Fig.~\ref{Figure2}. These novel features are,
in principle, mathematically \textit{impossible} for solutions of
ODEs. Therefore, regardless of the smallness of the time interval
over which past history must be prescribed, we can expect qualitatively
new features to emerge from the application of this new model to molecular
dynamics.

It is possible to conceptualise the electrodynamic $n$ body problem
in an alternative way, using fields. Since this has often been done
in the past, we clarify how the field picture relates to our approach.
To determine the field acting on a given particle, we must determine
the total field generated by all other particles. To do this, it is
necessary to solve Maxwell's equations. To solve these partial differential
equations (PDEs), it is necessary to provide initial (Cauchy) data
by prescribing the electromagnetic field over all space at one instant
of time (i.e., over a Cauchy hypersurface). Assuming retarded Lienard-Wiechert
potentials, the electromagnetic field due to a system of \textit{n}-particles,
at any instant of time, depends upon the \textit{past} motions of
those particles \cite{ckrem2bp}. Hence, prescribing this Cauchy data
requires a knowledge of the \textit{entire} past history of the positions,
velocities, and accelerations of the $n$ particles producing the
fields (Fig.~3).

The field picture, therefore, only hides the dependence on the past,
made explicit in the particle picture. In the field approach to the
electrodynamic $n$-body problem, we are required to solve a \textit{coupled}
system of ordinary and partial differential equations (PDEs). To determine
the motion of one particle, we need to solve the Newtonian ODEs with
the Heaviside-Lorentz force due to the fields generated by other particles.
Those fields are determined from particle motions using Maxwell's
equations. Compared to this coupled system of ODEs + PDEs in the field
picture, it is, currently, computationally more convenient to solve
the FDEs of the particle picture for reasons already discussed
elsewhere in detail in \cite{ckrem2bp}. Which computational technique is used to solve the equations is not, of course, relevant to the model of memory that is being proposed here.

\begin{figure}[htb]
 \includegraphics[height=12pc]{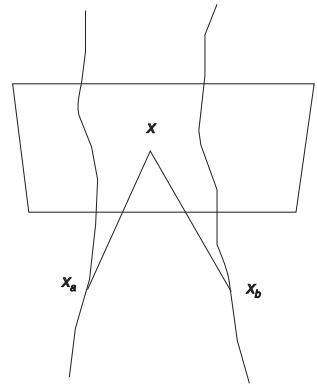}
\caption{\textbf{Relation of FDE method to ODE+PDE method.} Assuming only two
particles and retarded propagators, the electromagnetic fields at
any point $x$ on a Cauchy hypersurface relate to past particle motions
at points $x_{a}$ and $x_{b}$, where the backward null cone with
vertex at $x$ meets the world lines of the two particles $a$ and
$b$. As $x$ runs over the hypersurface, the points $x_{a}$ and
$x_{b}$ will, in general, cover the entire past world-lines of the
two particles. Thus, for solutions of the 2-body problem, the field
picture and the PDE+ODE method requests \textit{more} information
about the past than is practically needed by the particle picture
and the FDE method. }
\label{Figure3} 
\end{figure}

\section{The difference}

\begin{figure}[htb]
 \includegraphics[width=18pc]{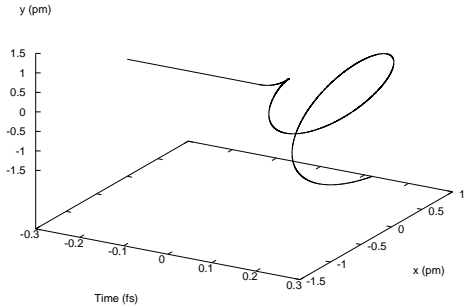}

\caption{\textbf{Difference between FDE and ODE solution for the 2-particle
problem.} For the case of the classical hydrogen atom, the figure
plots the time evolution of the difference between the electron orbit
with the full electrodynamic force and its orbit on the Coulomb force.
The zero difference part corresponds to the past history, prescribed
using the Coulomb orbit.}

\label{Figure4} 
\end{figure}

To ascertain the exact difference made by the use of the
full electrodynamic force in place of the electrostatic force, we have computed \cite{ckrem2bp}
the first numerical solutions of the retarded FDEs of the 2-body problem,
in a realistic physical context, using the full electrodynamic force,
but neglecting radiation damping. In the case of the classical hydrogen
atom, the solution with the full electrodynamic force (but without
radiative damping) differs from the Coulomb orbit, prescribed as past
data. This is summarily shown in Fig.~\ref{Figure4}. The
difference relates to an unexpected `delay torque' \cite{ckrem2bp} which arises 
because the full electrodynamic force depends upon the
past motion of the other particle. It is impossible for any central
force (like the electrostatic force) to have such a torque. Thus,
the electrodynamic interaction of two charged particles involves complexities
that cannot be captured by the simple Coulomb potential.

This suggests that a far richer variety of behaviour can be modeled
if the MD force field is reformulated by replacing the electrostatic
force in (1) by the full electrodynamic force (2).

What are the computational costs of solving FDEs, instead of the usual
ODEs? Stiffness considerations can be set aside since they apply equally to FDEs and ODEs. The relative increase in complexity comes from the calculation of the electrodynamic force (2) instead of the Coulomb force in (1). However, once the retarded time $t_r$ is known, this only adds a constant number of floating point operations per pair of interacting sites.  

Thus, the basic increase in complexity comes from having to solve a nonlinear equation to determine $t_r$ for each (ordered) pair. This is not as bad as it sounds, since, between time steps, each value of $t_r$ would be expected to change by only a small amount. So, with the previous value of $t_r$ as the starting guess, we can expect quick convergence in one or two steps. Thus, for practical purposes, the force calculation would asymptotically remain at most $O(n^2)$, though with a different constant. Further reduction of complexity by means of a cutoff is considered below.

Memory requirements would also increase, since some part of the past history of each interacting particle/site would have to be kept in memory to avoid excessive swapping and interpolation. Exactly how much of the past history needs to be retained in memory depends upon the specific algorithm used to solve the FDEs, and whether or not it permits step sizes larger than the interval of retardation. 

On the whole, it is reasonable to expect that, for most existing MD computer programs, the resulting increase
in computational complexity can be handled with existing computers---an exceptional case is \textsc{namd2} \cite{namd2}, 
and the parallel versions of \textsc{amber}.
While the solution of ODEs may be parallelised with reasonable efficiency   \cite{ckrdopri},
the history dependence of FDEs may be expected, by Amdahl's law, to
pose a serious challenge to parallel computing. However, Amdahl's law restricts only parallelised time-stepping. The force calculation, which is the compute-intensive part, can still be done in parallel.

The complexity can be reduced, as in the Coulomb case, by applying a long-range cutoff. This would reduce the memory requirements as well, though it is not so clear in the present context that this is necessarily desirable. The range of the full force, however, is larger, especially if we take into account the radiation damping. Until now, it was impossible to take into account the effects of radiation damping in a many-body problem, due to
various long-standing difficulties like preacceleration \cite{dirac,plass}. 
However, these difficulties were recently addressed in a satisfactory way, using FDEs  \cite{patomu}. Theoretically, or with a long-term outlook, it is a significant advantage of this proposal that the effects of radiation damping can also be included, if so desired. Inclusion of radiation damping would, however, add to the complexity by introducing a new source of stiffness in the problem.

Finally, we note that FDEs have long been linked to quantum mechanics, on the structured-time interpretation of quantum mechanics \cite{ckrtitcon}. A more detailed examination of exactly how the FDE approach relates to quantum mechanics in the context of MD will be considered in a subsequent paper.

\section{Conclusions}

It is desirable to reformulate MD to use the full electrodynamic force, which is more accurate than the Coulomb force.  This involves solving FDEs,  which is currently computationally feasible. This reformulation allows an exploration of  a qualitatively richer set of ways in which biological macromolecules can interact at long range, and may help to understand the possible dynamical origin of biological memory.

\bibliography{references}

\end{document}